\documentclass[fleqn,usenatbib]{mnras}

\usepackage{subcaption}
\usepackage{amsmath}	
\usepackage{mathptmx}
\usepackage{txfonts}

\usepackage[T1]{fontenc}

\DeclareRobustCommand{\VAN}[3]{#2}
\let\VANthebibliography\thebibliography
\def\thebibliography{\DeclareRobustCommand{\VAN}[3]{##3}\VANthebibliography}

\usepackage{graphicx}	

\title[Can Rotation Solve the Hubble Puzzle?]{Can Rotation Solve the Hubble Puzzle?}

\author[Balázs Endre Szigeti et al.]{
Balázs Endre Szigeti,$^{1,2}$
István Szapudi,$^{3}$\thanks{MTA Guest Professor 2023, Email: istvan@hawai.edu}
Imre Ferenc Barna $^{2}$
and  Gergely Gábor Barnaföldi$^{2}$
\\
$^{1}$Eötvös Loránd University, Institute of Physics, 11/A Pázmány Péter Stny, Budapest H-1117, Hungary\\
$^{2}$HUN-REN Wigner Research Centre for Physics, P.O. Box 49, Budapest H-1525, Hungary\\
$^{3}$Institute for Astronomy,  University of Hawaii, 2680 Woodlawn Drive, Honolulu, Hawaii,  96822,  USA
}

\date{Accepted XXX. Received YYY; in original form ZZZ}

\pubyear{\the\year{}}

\begin{document}
\label{firstpage}
\pagerange{\pageref{firstpage}--\pageref{lastpage}}
\maketitle

\begin{abstract}
    The discrepancy between low and high redshift Hubble constant $H_0$ measurements is the highest significance tension within the concordance $\Lambda$CDM paradigm. If not due to unknown systematics, the Hubble puzzle suggests a lack of understanding of the universe's expansion history despite the otherwise spectacular success of the theory. We show that a Gödel inspired slowly rotating dark-fluid variant of the concordance model resolves this tension with an angular velocity today $\omega_0 \simeq 2\times 10^{-3}$~Gyr\textsuperscript{-1}. Curiously, this is close to the maximal rotation, avoiding closed time-like loops with a tangential velocity less than the speed of light at the horizon.
\end{abstract}




\section{Introduction}\label{sec1}

The Hubble tension, the inconsistency of the late and early time measurements of the universe's expansion rate, emerges as the most significant chink in the otherwise shiny armour of the concordance $\Lambda$CDM model (see, e.g. for reviews by~\cite{verde2019,DiValentino2021,Kamionkowski2023}. The discrepancy has been established in a wide range of data sets and reached a $5\sigma$ significance between cepheid-calibrated local supernovae and cosmic microwave background (CMB) measurements \citep[for counterpoint and calibration uncertainties see][]{Freedman2024}.  

 The CMB constraints at recombination are indirect: they assume an expansion history governed by the $\Lambda$CDM model. The latest analyses of Planck CMB maps imply a Hubble constant $H_{\mathrm{CMB}} = 67.4 \pm 0.5 \>  \mathrm{km} \mathrm{s}^{-1} \mathrm{Mpc}^{-1}$~\citep{Planck2018}.

Type Ia supernovae directly constrain the late-time (local) expansion rate. In a definitive study of~\cite{Riess2022} used the Hubble Space Telescope (HST) to observe Cepheid variables in the host galaxies of 42 Type Ia supernovae (SNe Ia) crucial for calibrating the local Hubble constant ($H_{\mathrm{SNe}}$).
They utilized all suitable SNe Ia discovered at redshift $z \leq 0.01$ over the past four decades, significantly expanding the sample size with observations from over 1000 HST orbits.
They performed geometric calibration of Cepheids using Gaia EDR3 parallaxes, masers in NGC 4258, and detached eclipsing binaries in the Large Magellanic Cloud. Their baseline result is $H_{\mathrm{SNe}} = 73.04 \pm 1.07 \>\mathrm{km} \mathrm{s}^{-1} \mathrm{Mpc}^{-1} $, with systematic uncertainties, closely aligned with the median of various analysis variants. Notably, they found a significant 5$\sigma$ discrepancy with the Planck CMB analysis. 

A burst of activity resulted, including alternative (tip of the red giant branch) calibrations by~\cite{Freedman2019}, and extensions or modifications of $\Lambda$CDM by~\cite{DiValentinoa2020}, such as massive neutrino or WIMP models of~\cite{Knox2015}, dark photon of~\cite{Aboubrahim2022}, and an extended dark sector by~\cite{DiValentinob}. Next, we propose rotating space-time as a novel solution. 
G\"odel \cite{Godel1949} introduced a rotating universe followed by~\cite{Heckmann1955, Heckmann1956, Heckmann1955b} and~\cite{Heckmann1961} later~\cite{Silk1966} and~\cite{Hawking1969}. Visualization of the Gödel's universe is made by~\cite{Buser_2013}. While anisotropies in a variety of Bianchi models with large vector perturbations corresponding to rotation are tightly constrained from Planck CMB data by~\cite{Saadeh2016}, generalizations of the Gödel model by~\cite{Obukhov2000} with global rotation are still viable and free of the pathologies of the original. This paper considers a Newtonian approximation of these models in the context of the Hubble anomalies. 

All objects within our universe rotate, including planets, stars, solar systems, galaxies, and galaxy clusters. Moreover, black holes, spherically symmetric objects with horizons, display near maximal rotation as presented by~\cite{Daly2019}. The idea that everything revolves ($\pi \alpha \nu \tau \alpha $ $ \kappa \upsilon \kappa \lambda o \upsilon \tau \alpha \iota$)~\footnote{\emph{Panta kykloutai} paraphrasing the ancient Greek philosopher Heraclitus} naturally extends to the whole universe, as hinted by recent claims of anisotropic Hubble expansion in X-ray observations by~\cite{Migkas2024}.
Furthermore, a plausible syllogism is that the universe has near-maximal rotation, motivated by cosmologies where the universe is the interior of a black hole~\citep{PATHRIA1972}. There are many proposed solutions to the Hubble Puzzle \citep[e.g., ][]{HubbleRev2021} and any modification of the standard model expansion and growth history has to consider the entire concordance model \citep[e.g.,][]{HubbleHunter2020}. Nevertheless, as we show later, the average rotation effect has a similar functional form to that of dark photons \citep{Fabbrichesi2021,Aboubrahim2022}, one of the promising contenders \citep{DarkPhoton2022} for solving the Hubble Puzzle. Therefore, exploring how a rotating model would affect the Hubble constant is worthwhile. In the next section, we outline our methodology, section~3 presents the results, while the last section contains our conclusions.

\section{Methodology}\label{sec2}

The expansion history of Newtonian cosmological simulations is in precise agreement with Friedmann models \citep[e.g.,][]{Racz2018}. Thus, we expect that the classical framework is sufficient for an initial estimate of rotational effects on the Hubble constant; we leave general relativistic considerations for future work.

\begin{figure}
\centering
\includegraphics[width=0.45\textwidth]{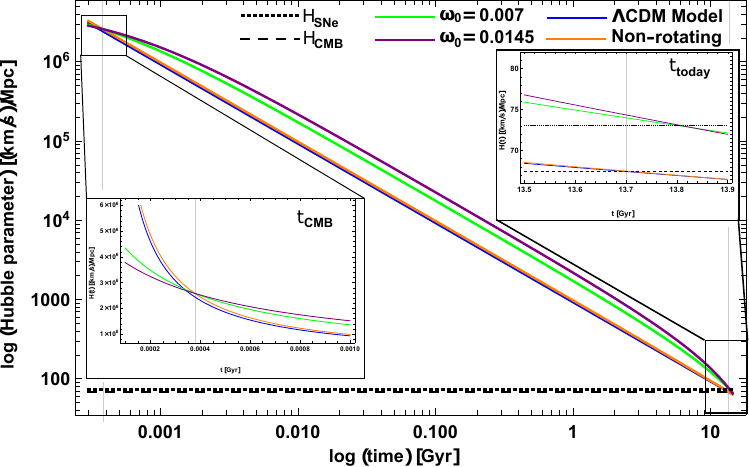}
\caption{Time evolution (log-log scale) of the Hubble parameter for non-rotating (analytical) and rotating (numerical) solutions at different $\omega_0$ rotation parameter values as of today. Small figures show the evolution (normal scale) at the decoupling period $t_{\rm CMB}$ and today $t_0 = t_{\rm today}$.} 
\label{fig::HubbleTimeEvolution1}
\end{figure}

Describing the evolution of the Hubble parameter in Newtonian non-rotating and rotating universe models is still challenging. The Sedov\,--\,von Neumann\,--\,Taylor blast wave models inspired us to 
construct a non-relativistic dark fluid model. We apply a non-linear partial differential equation system describing a non-viscous, non-relativistic, and self-gravitating fluid with zero thermal conductivity (Euler\,--\,Poisson system) and solve it with a time-dependent Sedov-type self-similar \emph{ansatz}. This analytic approach incorporates various scaling mechanisms and describes different time decay scenarios of~\cite{Taylor1950}. The resulting dynamical model is consistent with direct solutions from the Friedmann equations by~\cite{Szigeti2023}. We generalize our method of intermediate asymptotic analysis of the hydrodynamical description for the rotating dark-fluid universe to investigate the effect of rotation on
the Hubble-constant anomaly. Our partial differential equations read as follows: 
\begin{subequations}
\begin{align}
    \partial_t \rho + \mathrm{div} (\rho \boldsymbol{u}) &= 0 \label{eq:01A} \ , \\
    \partial_t (\rho \boldsymbol{u}) + \mathrm{div} ( \rho \boldsymbol{u} \otimes \boldsymbol{u} )  &= - \nabla P(\rho) - \rho \nabla \Phi + \rho \boldsymbol{g^{\ast}} \ , \label{eq:01B}\\
    \nabla^2 \Phi &= 4 \pi G \rho ,   \label{eq:01C} 
\end{align}
\label{eq:01}
\end{subequations}
where $\rho,{\bf{u}}, P, \Phi, {\bf{g}}$ are  
the fluid density, the fluid velocity vector, the pressure, the
gravitation potential and the external force, respectively. 
This system has been investigated previously by~\cite{Deng2003,Wong2020}. \cite{Goldreich1980} 
studied the homologously collapsing stellar cores 
with an adiabatic exponent. Later 
\cite{Yuen2009} gave analytically periodic solutions to the 3-dimensional Euler\,--\, Poisson equations of gaseous stars with negative cosmological constants, commonly used to describe a dark-fluid system.  In this model, dark matter and dark energy are two different aspects of the same substance, the "dark fluid"  \cite{Farnes2018}. As illustrated in Fig.~\ref{fig::HubbleTimeEvolution1}, the self-similar solution using the specific dark fluid equation of state yields results consistent with $\Lambda$CDM within the relevant time range.

\begin{description}
    \item[Spherical symmetry:] Initially, we assumed an ideal fluid with spherical symmetry. Therefore, the multi-dimensional partial differential equation system reduces to a one-dimensional, radius-dependent ordinary system. We assume a linear equation of state (EoS) by~\cite{Horedt2004}, $P (\rho) = w \rho$ with the effective $w$ asymptotically tending to -1 at $t_{\infty}$. The resulting dark-fluid models describe a mixture of dark matter and dark energy in a non-rotating, expanding universe ~\citep{Szigeti2023}, as long as we neglect dark matter fluctuations ~\citep[see extensive studies by][]{GUO2007326}. The Euler-Poisson equation in the spherical limit is the following:
\begin{align}
    \partial_t \rho + (\partial_r \rho) u + (\partial_r u) \rho + \dfrac{2u\rho}{r} &= 0 \label{eq:02A} \ , \\
    \partial_t u + (u \partial_r) u &= - \dfrac{1}{\rho} \partial_r P  - \partial_r \Phi(r) + g^{\ast}\label{eq:02B} \ , \\
    \dfrac{1}{r^2} \dfrac{d}{dr}\left(r^2 \partial_r \Phi \right)  & = 4 \pi \rho\label{eq:02C} \ ,
\end{align}
where the $u(r,t)$ radial flow velocity.
Equations~\eqref{eq:01A}-\eqref{eq:01C} are reduced to Eqs.~\eqref{eq:02A}-\eqref{eq:02C} due to the high similarity of the system as demonstrated by \cite{Sedov1959}.

\item[Cylindrical symmetry:] If the rotation becomes significant enough, assuming spherical symmetry is no longer adequate. We assumed that the system is fully symmetric in the $z$ direction. Thus, we extend our previous analyses of spherical flows to the Euler-Poisson equation in a cylindrical coordinate system. 
\begin{align}
    \partial_t \rho + (\partial_r \rho) u + (\partial_r u) \rho + \dfrac{u\rho}{r} &= 0 \label{eq:03A} \ , \\
    \partial_t u + (u \partial_r) u &= - \dfrac{1}{\rho} \partial_r P  - \partial_r \Phi(r) + g^{\ast}\label{eq:03B} \ , \\
    \dfrac{1}{r} \dfrac{d}{dr}\left( r \partial_r \Phi \right)  & = 4 \pi \rho\label{eq:03C} \ .
\end{align}
\cite{Holden2024} studied self-similar solutions for the infinite cylindrical collapse.  
Self-similar exponents similar to the spherical case exist orthogonal to the $z$ axis. The $u(r,t)$ again has a similar meaning as defined in the spherical symmetric case.
   
\end{description}
We solved both sets of the equations by using the Sedov\,--\, Taylor ansatz for the velocity field $u(r,t)$, the density $\rho(r,t)$, and the gravitational potential density field $\Phi(r,t)$ in both system. For spherical symmetry we applied the $u(r,t)  =  t^{-\alpha} f(\eta)$, $\rho(r,t)  = t^{-\gamma} g (\eta)$, and $\Phi(r,t) = t^{-\delta} h(\eta)$ notation, where $f(\eta)$, $g(\eta)$, and $h(\eta)$ are the shape functions of the reduced ordinary differential equation system with the reduced variable, $\eta = r/t^{\beta}$. The analogous shape functions for cylindrical symmetry depend on the $z$ coordinate: $u(z,t)  =  t^{-\alpha} k(\xi)$, $\rho(z,t)  = t^{-\gamma} l (\xi)$ and $\Phi(z,t) = t^{-\gamma} h(\xi)$, where $\xi = z/t^{\beta}$.

The real parameters $\alpha, \beta, \gamma$, and $\delta$ are the self-similar exponents responsible for the solution's temporal decay and spreading. The numerical values of the exponents: $ \alpha = 0$, $ \beta = 1$, $ \gamma = 2$ and $ \delta = 0 $. We assumed similarly to the spherical case for the cylindrical case that the dynamical variables depend only on the $\eta = r/t^{\beta}$ variables. We performed all calculations in physical (non-expanding ) rotating coordinates. The scale factor is the solution of the differential equation of $\Dot{a}(t) = u(r(t),t)$ (for details, see technical \cite{Szigeti2023}), i.e. the approximately uniform streaming of the particles in physical coordinates represents the expansion of the universe. We restricted to expanding solutions of Eqs.~\eqref{eq:02A}-\eqref{eq:03C} with $r \propto a(t)$. Consequently, the Hubble parameter is expressed as usual $H(t) = \Dot{a}/a$. 

We solved the two systems under equivalent initial conditions in $\eta$. We have seen that the two systems exhibit analogous behaviour in regions where the intermediate-asymptotic behaviour yields a valid solution. Thus, we will use the spherically symmetric limit in our following analyses. (See in Fig.~\ref{fig::HubbleTimeEvolution1}).
We assumed spherical symmetry for our self-similar solutions of the Euler\,--\,Poisson system. This is justifiable far from the boundary, and it is numerically more stable \cite{PENG20126370,Yuen2009}.

The same mathematical framework describes the Hubble parameter for the non-rotating ($\boldsymbol{g^{\ast}}=0$) and the rotating model ($\boldsymbol{g^{\ast}} \neq 0$) by setting initial conditions for the density, the velocity and the gravitational field: $u(\eta_{\textrm{IC}}) = 0.5$ and $\rho(\eta_{\textrm{IC}}) = 0.01$ $ \mathrm{l}_{ \textrm P}^{-1}/\mathrm{m}^{3}$ in geometrized units ($G = c = 1)$ at $t_{\textrm{CMB}} = t_{\textrm{IC}} = 380$~kyr to determine the shape functions \footnote{in SI: $\sim 10^{-15} \mathrm{kg}/\mathrm{m}^{3}$  and $h(\eta_{\textrm{IC}}) = 10^{-12}$}. Similary, in case of cylindrical symmetric system the $k(\xi_{ \textrm{IC}}) = 0.5$ and $l(\xi) = 0.01$. These initial conditions are consistent with \cite{Planck2018}, and, as we have shown, they reproduce the no-rotation Friedman solutions. These initial conditions are consistent with the Planck initial conditions at the decoupling period.
We can add an effective rotational term to Equation~\eqref{eq:01B},
\begin{equation}
    \boldsymbol{g^{\ast}} = 2r \omega^2(t) \sin(\theta)  ,
    \label{eq::centrifugal}
\end{equation}
where the angular velocity is $\omega(t) = (\omega_0/t_0) / t$ and $\theta$ is the polar angle. The effective centrifugal force is given in a non-inertial rotating frame. The Coriolis force vanishes since the velocity from expansion is always perpendicular to the rotation axis. 
Slow rotation can still be consistent with present observations. A slight global rotation still preserves a uniform CMB \citep{Obukhov2000,Saadeh2016,McEwen2013}. Soon, such a rotation might be constrained by comparing the local inertial frame with that of quasars in ~\cite{Szapudi2021}.
 
\section{Results and Discussion}\label{sec4}

We numerically calculate the evolution equations following~\cite{Szigeti2023}, transforming the equations into the co-moving frame and applying the constraints detailed in Section (5). 
Fig.~\ref{fig::HubbleTimeEvolution1} shows the time dependence of the Hubble parameter for various angular frequencies from recombination until the present. Different initial rotations result in different $H_0$ values today, but all solutions converge to zero at the asymptotic limit, $t \to \infty$. The evolution of the Hubble parameter with initial rotation in $\omega (t) \to 0$ limit approaches the non-rotating model. However, the limit is extrapolated due to numerical instabilities for extremely small $\omega_0$ values today. As a test, the black solid line displays the standard $\Lambda$CDM result~\footnote{The curve is evaluated by using the matter-dominated scale factor $a(t) = \left( 3 H_{\rm CBM;0} t/2 \right)^{2/3}\Omega_m^{1/3}$, with the value of $\Omega_m=0.3089$.}, in perfect agreement with our formalism (the orange solid line labelled non-rotating). The non-rotating self-similar solution is consistent between $t_{\rm CMB}$ and $t_{\rm today}$~\footnote{An appropriate choice of initial conditions for the scale factor in Eq.~(25) in the work of~\cite{Szigeti2023} can be reduced, due to $u_2 \sim \mathcal{O}(10^{-4})$}.
\begin{figure}
\centering
\includegraphics[width=0.45\textwidth]{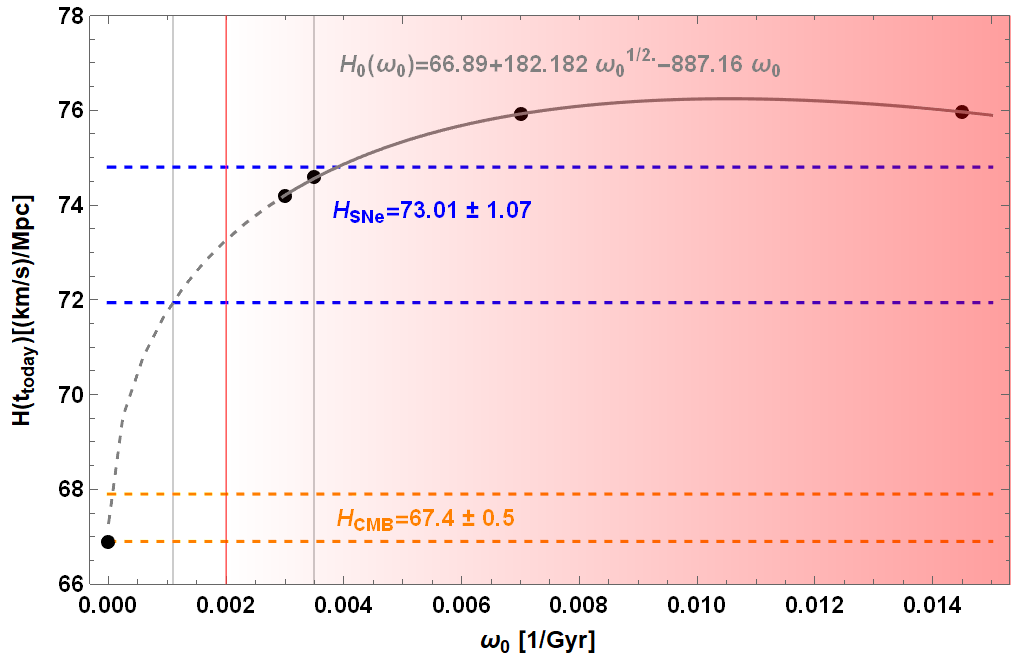}
\caption{The predicted $H_0$ values from non-rotating (analytical) and rotating (numerical) models evaluated at different $\omega_0$ values today. The grey line interpolates the calculated values (markers). The dashed line extrapolates the $\omega_0 \to 0$ case. The orange ($H_{\mathrm{CMB}}$)~\citep{Planck2018} and the blue ($H_{\mathrm{SNe}}$)~\citep{Riess2022} lines correspond to measurements with $2\sigma$ uncertainty ranges. Red shading approximates the prohibited region exceeding maximal rotation.} 
\label{fig::HubbleTimeEvolution2}
\end{figure}
Fig.~\ref{fig::HubbleTimeEvolution2} displays the Hubble constant ($H_0$) for the non-rotating (analytical) and various slowly-rotating (numerical) cases as a function of the rotational parameter $\omega_0$. The solid grey curve represents interpolation to the numerical calculations (markers), while the dashed curve is an extrapolation for $\omega_0 \to 0$. Numerical extrapolation for the Hubble constant, $H_0$ with $\omega_0 = 0.002^{+0.001}_{-0.0009}$~Gyr$^{-1}$ predicts a value today comparable to the measured by $H_{\mathrm{SNe}}$. The present day $\omega_0$ rotation corresponds to an initial, $\omega (t_{\rm CMB})=  3.54^{+1.3}_{-1.2}$~Myr$^{-1}$, where $H_{\rm CMB}$ is measured  at $t_{\mathrm{CMB}}=380$~kyr. 
\begin{figure}
    \centering
    \begin{subfigure}[b]{0.22\textwidth}
        \centering
        \includegraphics[width=\textwidth]{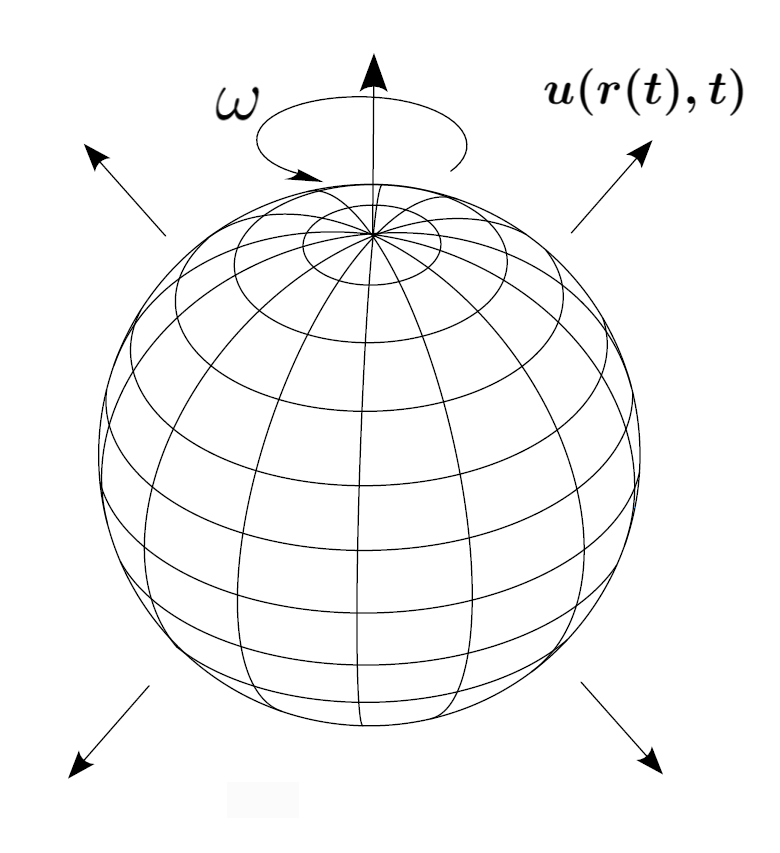}
    \end{subfigure}
    \hfill
    \begin{subfigure}[b]{0.22\textwidth}
        \centering
        \includegraphics[width=\textwidth]{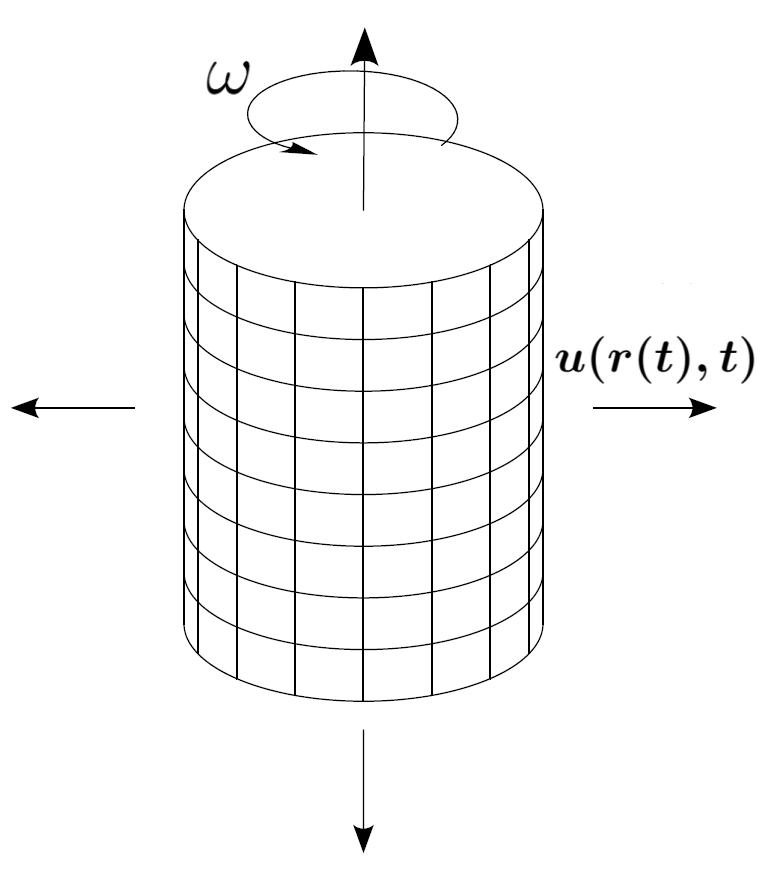}
    \end{subfigure}
    \caption{Schematic view of the rotating (non-expanding) spherical and cylindrical physical coordinate systems, where the outflow of the particles, $u(r(t),t)$ determines the expansion rate through $\dot{a}(t)= u(r(t),t)$; $\omega$ refers to the angular velocity. Even though $u(r(t),t)$ has a formal dependence on $r$, our solutions produce an outflow uniform enough that it is well described by a single expansion rate, $a(t)$.}
    \label{wrap-fig:1}
\end{figure}
In Fig.~\ref{wrap-fig:1}, we illustrate the rotating universe. Its angular rotation parameter, $\Vec{\boldsymbol{\omega}}(t)$, is approximately  
\begin{equation}
    |\Vec{\boldsymbol{\omega}}(t)| = \omega_0 a^{-2}(t)  
\end{equation}
from angular momentum conservation during matter domination, consistently with Eq.~\eqref{eq::centrifugal}. Next, we estimate the maximal rotation of a dark-matter-filled universe and compare it with $\omega_0 \simeq 0.002$ solving the Hubble tension. We require that the speeds remain below the speed of light within the observable horizon, hence $\omega \lesssim H$, during the universe's entire history. Taking $H(a) \sim a^{-3/2}$, we limit $\omega_0$ today as,
\begin{equation}
    \omega_0 \lesssim H_0 a^{1/2}(t_{\mathrm{eq}}) \simeq 0.002\ \mathrm{Gyr}^{-1},
\end{equation} 
where $t_{\mathrm{eq}}$ is the time of matter-radiation equality. Note that since at earlier times $H(a)\simeq 1/a^2$ and
$\omega(a) \simeq 1/a$, the above condition is satisfactory for the entire evolution of the universe.
Most remarkably, \emph{the allowed maximal rotation is approximately the same as the one required to solve the Hubble Puzzle.} Our simplified argument neglected any late effects of Dark Energy on angular momentum, but there should be a reasonable estimate for our calculation. Our results are consistent with~\cite{Heckmann1955, Heckmann1956, Heckmann1955b} and~\cite{Heckmann1961}, despite the differences in techniques and their original motivation of removing the initial singularity at the Big Bang. The required a \emph{minimal} rotation, $\omega \simeq 0.03~\mathrm{Gyr}^{-1}$, is an order of magnitude larger than the \emph{maximal} rotation avoiding closed time-like loops within the horizon.

\section{Conclusion}\label{sec6}

We analyze the time evolution of the Hubble parameter within the Euler\,--\,Poisson model with a self-similar time-dependent Sedov-type scaling for a linearized dark-fluid equation of state. This model is consistent with a Newton-Friedmann cosmology when the angular momentum is zero and facilitates the analysis of cosmologies with slow rotation.

We found that an angular speed near the maximal rotation $\omega_0 \lesssim 0.002\ \mathrm{Gyr}^{-1}$ today predicts a Hubble constant consistent with local measurements when starting from an expansion rate consistent with the CMB.
Extrapolation to the initial rotation of the early universe gives the values of $\omega (t_{\rm CMB}) \approx 3.54^{+1.3}_{-1.2}$Myr$^{-1}$ for the time of the origin of the cosmic microwave background. These tantalizing initial results have the caveat that we only focused on the Hubble constant. Further investigations contrasting the rotating model against the entire intertwined network of the concordance model observations, confirmation and development of numerical models using rotating cosmological $N$-body simulations \footnote{P\'al et al. 2025 in prep.}, and extension for a general relativistic treatment are left for future work.

\section{Acknowledgement}\label{sec7}

We acknowledge the financial support by the Hungarian National Research, Development and Innovation Office (NKFIH) under Contract No. OTKA K135515, 2021-4.1.2-NEMZ\_KI-2024-00031, 2024-1.2.5-TÉT-2024-00022 and Wigner Scientific Computing Laboratory (WSCLAB, the former Wigner GPU Laboratory) IS acknowledges support by the National Science Foundation under grant no. 2206844 and thanks the hospitality of the MTA-CSFK Lend\"ulet "Momentum" Large-Scale Structure (LSS) Research Group at Konkoly Observatory supported by a \emph{Lend\"ulet} excellence grant by the Hungarian Academy of Sciences (MTA).

\section{Data Availability}
No new data were generated or analysed in support of this research. The software code developed for this article will be available upon reasonable request.



\bibliographystyle{mnras}
\bibliography{example} 





\bsp	
\label{lastpage}
\end{document}